\pdfoutput=1%
\documentclass[a4paper,british,citeautoscript,floatfix,pdftex,superscriptaddress,twoside,%
pra,aps,%
reprint,final%
]{revtex4-2}%
\usepackage{amsfonts,amsmath,amssymb}
\usepackage{hyperref}
\usepackage[T1]{fontenc}
\usepackage{graphicx}%
\usepackage{xspace, textcomp}
\usepackage[utf8]{inputenc}
\usepackage{xr}
\usepackage{commath, etoolbox, ifthen, letltxmacro, microtype, xcolor, xparse}

\graphicspath{{./fig/}} 

\renewcommand{\figurename}{Figure}

\makeatletter
\renewcommand{\fnum@figure}{{\normalfont\bfseries\figurename~\thefigure}}
\renewcommand{\@caption@fignum@sep}{\textbf{ | }}
\makeatother

\newcommand*{\methods}{\hyperref[sec:methods]{Methods}\xspace}%
\newcommand*{\supsec}[1]{Supplementary Note #1\xspace}

\newcommand*{\ind}{\ensuremath{\text{C}_8\text{H}_7\text{N}}\xspace}
\newcommand*{\indw}{\ensuremath{\ind\text{--}\HHO}\xspace}
\newcommand*{\indolew}{\ensuremath{\text{indole}(\HHO)}\xspace}
\newcommand*{\indolewp}{\ensuremath{\indolew^+}\xspace}
\newcommand*{\indolewn}[1]{\ensuremath{\text{indole}(\HHO)_{#1}}\xspace}
\newcommand*{\indole}{\ensuremath{\text{indole}}\xspace}
\newcommand*{\indolep}{\ensuremath{\text{indole}^+}\xspace}
\newcommand*{\HHO}{\ensuremath{\text{H}_2\text{O}}\xspace}
\newcommand*{\HHOp}{\ensuremath{\text{H}_2\text{O}^+}\xspace}

\newcommand*{\HHHOp}{\ensuremath{\text{H}_3\text{O}^+}\xspace}
\newcommand*{\HHOd}{\ensuremath{(\HHO)_2}\xspace}

\newcommand*{\indolean}[1]{\ensuremath{\text{indole}(\NHHH)_{#1}}\xspace}
\newcommand*{\NHHH}{\ensuremath{\text{NH}_3}\xspace}
\newcommand*{\pps}{\ensuremath{\pi\pi^*}\xspace}
\newcommand*{\pss}{\ensuremath{\pi\sigma^*}\xspace}
\newcommand*{\rhos}[2]{\ensuremath{\rho_{#1#2}}\xspace}
\newcommand*{\rhosdot}[2]{\ensuremath{\dot{\rho_{#1#2}}}}
\newcommand*{\eg}{e.\,g.}%
\newcommand*{\ie}{i.\,e.}%
\newcommand*{\ordsim}{\mathord{\sim}}
\newcommand*{\larger}{\mathord{>}}
\newcommand*{\Ei}{\ensuremath{\xspace{}E_i}\xspace}
\newcommand*{\Wpcmcm}{\ensuremath{\text{W}/\text{cm}^2}\xspace}
\newcommand*{\um}{\ensuremath{\text{\textmu{m}}}\xspace}
\newcommand*{\celsius}[1]{\ensuremath{\xspace#1\,^\circ{}\text{C}}\xspace}
\newcommand*{\invcm}{\ensuremath{\text{cm}^{-1}}\xspace}
\newcommand*{\range}[2]{#1\ensuremath{\dots}#2}

\AtBeginDocument{%
	\LetLtxMacro\autoreforig\autoref%
	\RenewDocumentCommand{\autoref}{som}{%
		\IfBooleanF{#1}{%
			\hyperref[#3]%
		}{%
			\autoreforig*{#3}\IfValueT{#2}{\nobreak}\xspace\IfValueT{#2}{#2}%
		}%
	}
}

\newcommand{\cfeldesy}{\affiliation{Center for Free-Electron Laser Science, Deutsches
      Elektronen-Synchrotron DESY, Notkestraße 85, 22607 Hamburg, Germany}}%
\newcommand{\uhhcui}{\affiliation{Center for Ultrafast Imaging, Universität Hamburg, Luruper
      Chaussee 149, 22761 Hamburg, Germany}}%
\newcommand{\uhhphys}{\affiliation{Department of Physics, Universität Hamburg, Luruper Chaussee 149,
      22761 Hamburg, Germany}}%
\newcommand*{\uninijmegen}{\altaffiliation{Current address: Institute for Molecules and Materials,
      Radboud University, Heijendaalseweg 135, 6525 AJ Nijmegen, The Netherlands}}%
\newcommand{\jkemail}{\email[Email:~]{jochen.kuepper@cfel.de}}%
\newcommand{\cmiweb}{\homepage[website:~]{https://www.controlled-molecule-imaging.org}}%

\makeatletter%
\renewcommand*{\@fnsymbol}[1]{\ensuremath{\ifcase#1\or \mathsection\or *\or \|\or \ddagger\or **\or
      \mathparagraph\or \dagger\or \dagger\dagger \or \ddagger\ddagger \else\@ctrerr\fi}}%
\makeatother

\begin{document}
\title{Ultrafast light-induced dynamics in the microsolvated biomolecular indole chromophore with
   water}%
\author{Jolijn Onvlee}\uninijmegen\cfeldesy\uhhcui%
\author{Sebastian Trippel}\cfeldesy\uhhcui%
\author{Jochen Küpper$^\ast$}\jkemail\cmiweb\cfeldesy\uhhcui\uhhphys%
\date{\today}%
\begin{abstract}\noindent
   {\centering\textbf{ABSTRACT}\\[1ex]}
   Interactions between proteins and their solvent environment can be studied in a bottom-up
   approach using hydrogen-bonded chromophore-solvent clusters. The ultrafast dynamics following
   UV-light-induced electronic excitation of the chromophores, potential radiation damage, and their
   dependence on solvation are important open questions. The microsolvation effect is challenging to
   study due to the inherent mix of the produced gas-phase aggregates. We use the electrostatic
   deflector to spatially separate different molecular species in combination with pump-probe
   velocity-map-imaging experiments. We demonstrate that this powerful experimental approach reveals
   intimate details of the UV-induced dynamics in the near-UV-absorbing prototypical biomolecular
   indole-water system. We determine the time-dependent appearance of the different reaction
   products and disentangle the occurring ultrafast processes. This approach ensures that the
   reactants are well-known and that detailed characteristics of the specific reaction products are
   accessible -- paving the way for the complete chemical-reactivity experiment.
\end{abstract}
\maketitle%

It is a long-held dream of chemistry to follow chemical reactions in real
time~\cite{Polanyi:ACR28:119, Zewail:JPCA104:5660, Ischenko:CR117:11066}. Here, the observation of
the transition state~\cite{Polanyi:ACR28:119} and the recording of electronic and nuclear
motion~\cite{Blaga:Nature483:194, Calegari:Science346:336} during the making and breaking of bonds
are of special interest. These processes at the heart of chemistry occur on ultrafast attosecond
(as) to picosecond (ps) timescales. One of the fundamental challenges in their investigation is to
initiate the reactions effectively instantaneously on the timescale of the
dynamics~\cite{Zewail:JPCA104:5660, Nisoli:CR117:10760}. In the ultimate chemical-reaction-dynamics
experiment the reactants are well-defined and well-known, the characteristics of the products, \eg,
their yields, momenta, and structures, are precisely observed, and the intermediate electronic and
nuclear structures are precisely recorded with high spatial and temporal resolution.

In the ongoing quest for this ultimate experiment, scientists designed increasingly advanced
machines to prepare well-defined reactants~\cite{Meerakker:CR112:4828, Chang:IRPC34:557} and to
probe the characteristics of the products in high detail~\cite{Eppink:RSI68:3477}. These
ingredients already enabled extremely detailed studies of elementary chemical reactions involving
atoms and small molecules~\cite{Xie:Science368:767, Lin:Science300:966} including full details on
the quantum-state-correlations of reactants and products~\cite{Qiu:Science311:1440,
   Kirste:Science338:1060}, albeit without temporally resolving the chemical dynamics.

Alternatively, investigating photochemically-triggered reactions in pump-probe experiments with
ultrashort laser pulses enabled the direct study of chemical dynamics in real
time~\cite{Zewail:JPCA104:5660, Hertel:RPP69:1897}. Using intense ultrashort laser pulses to excite
strong electronic transitions with subsequent photochemical rearrangements and possibly breaking of
the molecules has two crucial advantages: it directly defines the starting time of the chemical
reaction and it also yields significantly higher densities of molecules undergoing the chemical
reaction. These dynamical molecular systems can then be probed using
ion-~\cite{Stapelfeldt:PRL74:3780, Ullrich:RPP66:1463} or electron-imaging
techniques~\cite{Baumert:PRL64:733, Holmegaard:NatPhys6:428}, high-harmonic-generation
spectroscopy~\cite{Itatani:Nature432:867, Vozzi:NatPhys7:822} and laser-induced electron
diffraction~\cite{Blaga:Nature483:194, Wolter:Science354:308, Meckel:Science320:1478}, or
x-ray~\cite{Barty:ARPC64:415, Kuepper:PRL112:083002} and electron
diffraction~\cite{Yang:Science361:64, Ischenko:CR117:11066}.

Severe limitations still exist for such time-resolved studies of bimolecular reaction systems. The
ultrafast dynamics of these complex reaction systems are studied through so-called
half-collisions~\cite{Zewail:Science242:1645}, in which a chemical bond between two molecules is
broken due to the laser excitation. Bimolecular aggregates, so-called molecular clusters, are
produced in a supersonic expansion. However, this yields a broad distribution of cluster sizes and
whereas the mean and width of the distribution can be tuned~\cite{Kappes:MolecularBeamsofClusters,
   Hertel:RPP69:1897}, it does not allow the production of samples of individual aggregates. In
combination with the necessarily broad bandwidth of ultrashort laser pulses, one always
simultaneously initiates and probes the dynamics of multiple molecular systems, with different
structures and sizes. This leads to overlapping signals from different chemical
reactions~\cite{Hertel:RPP69:1897}.

\begin{figure*}
   \includegraphics[width=\textwidth]{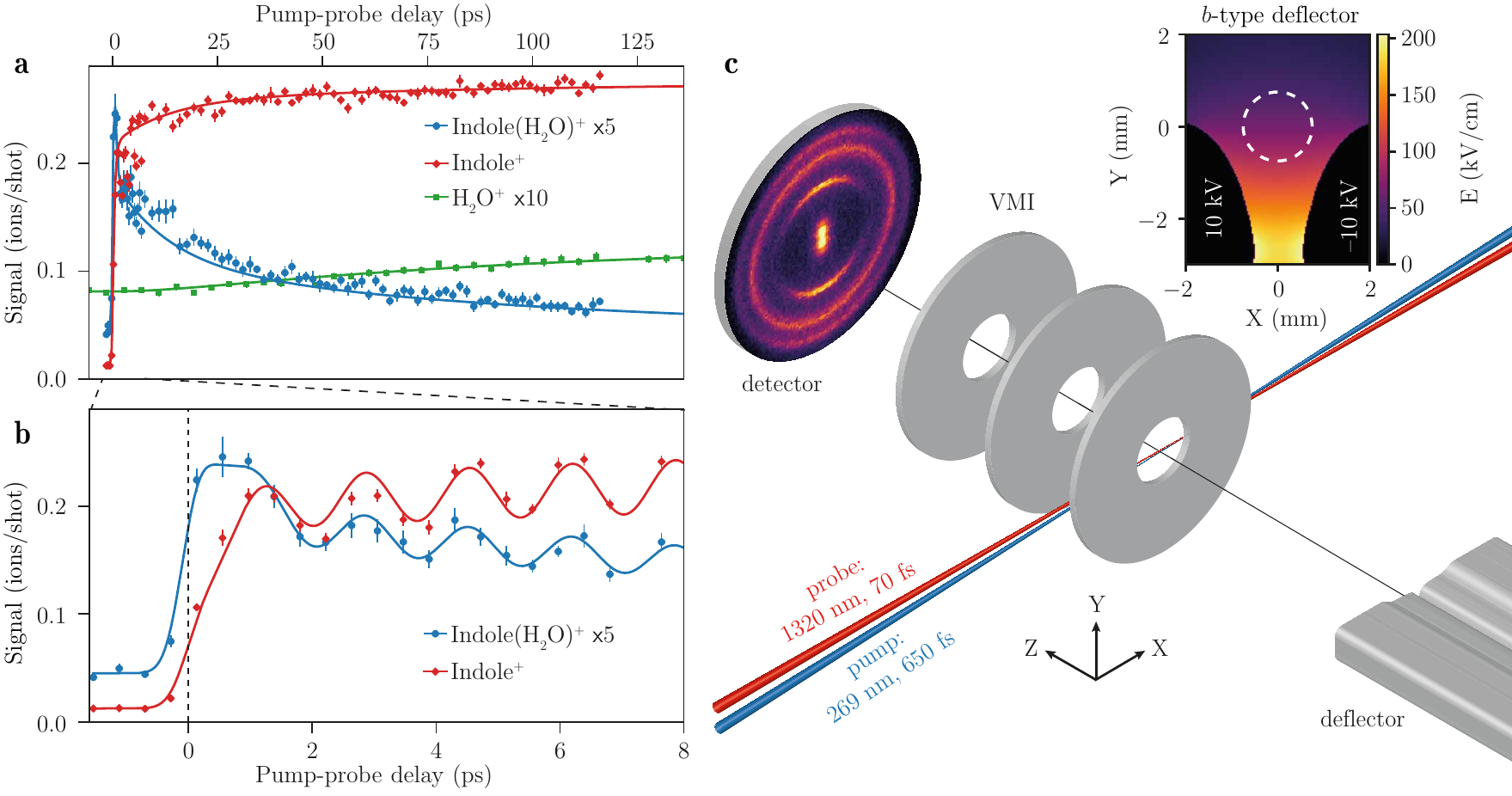}%
   \caption{\textbf{Time-dependent molecular signals and experimental setup.}
      \textbf{a}~Delay-dependent yields of (blue) \indolewp, (red) \indolep, and (green) \HHOp.
      Symbols represent the experimental data whereas lines represent the simulated data based on
      our reaction model. Statistical $1\sigma$ error estimates are given by vertical lines for all
      experimental data points. \textbf{b}~Delay-dependent \indolewp and \indolep signals for short
      delays. \textbf{c}~Illustration of the experimental setup, containing an electrostatic
      deflector and a velocity-map imaging (VMI) detector. A pump laser electronically excited
      \indolew, and a delayed probe laser ionised the reaction products. See \methods for details.
      The inset shows a cross section of the deflector: colours indicate the electric field strength
      $E$ and the white dashed circle indicates where the direct molecular beam passes through the
      deflector.}
   \label{fig:yields}
\end{figure*}

Small hydrogen-bonded aggregates of aromatic molecules or chromophores with polar solvent molecules
like ammonia (\NHHH) and water (\HHO) are important model systems for the interactions between
proteins and their solvent environment~\cite{Zwier:ARPC47:205, Sobolewski:PCCP4:1093}. For instance,
these interactions affect the folding and thereby directly the function of
proteins~\cite{Colombo:Science256:655}. Microsolvated biomolecular chromophores and their ultrafast
dynamics were extensively studied, but current knowledge of their dynamics is strongly limited due
to the problem of overlapping signals described above~\cite{Lippert:CPL376:40, Hertel:RPP69:1897}.

The indole-water (\indolew, \indw) complex, for instance, is of high relevance, as \indole is the
chromophore of the most
strongly near-ultraviolet (UV) absorbing common amino acid tryptophan~\cite{Sobolewski:CPL329:130}
and thus
proteins. It was shown that the environment strongly affects the photochemical properties of
tryptophan~\cite{Vivian:BiophysJ80:2093}. These characteristics were frequently used to investigate
the structure and dynamics of proteins~\cite{Pal:PNAS99:1763}. In addition, the UV-light-induced
dynamics of bare and microsolvated indole (\ind) was studied extensively both theoretically and
experimentally~\cite{Park:ACIE47:9496, Godfrey:PCCP17:25197, Sobolewski:PCCP4:1093,
   Korter:JPCA102:7211, Kuepper:PCCP12:4980, Lippert:CPL376:40, Conde:CP530:25}.

When hydrogen-bonded aggregates of chromophores and polar solvent molecules are irradiated by UV
light, they are typically electronically excited to one or several \pps
states~\cite{Sobolewski:JPCA111:11725}, in the case of indole most likely two, \ie, $^1L_b$ and
$^1L_a$~\cite{Brand:PCCP12:4968}. These states often possess a conical intersection with an
optically dark and dissociative \pss state that plays an important role in the photochemistry of
these species~\cite{Sobolewski:PCCP4:1093, Ashfold:Science1637:1640}. For \indole in this \pss state
it is expected that it ejects an electron into the aqueous environment leading to the formation of a
charge-separated state, or solvated electron~\cite{Sobolewski:CPL329:130}. For the \indolew
aggregate it was predicted that this electron-transfer process is followed by the transfer of a
proton, leading to a net hydrogen-transfer reaction from \indole to
\HHO~\cite{Sobolewski:PCCP4:1093}. However, so far this hydrogen transfer has not been observed
experimentally, although it was observed for the similar indole-ammonia (\indolean{n})
system~\cite{Lippert:PCCP6:2718}, which was ascribed to ammonia being a better hydrogen acceptor
than water~\cite{Sobolewski:PCCP4:1093}.

Previous studies on UV-excited \indolew clusters found dynamics occurring on multiple timescales of
\range{20}{100}~fs, \range{150}{500}~fs, and 14~ps~\cite{Lippert:CPL376:40, Conde:CP530:25}. These
were tentatively ascribed to an internal conversion between electronically excited states,
relaxation dynamics along the \pss state, and a coupling of the \pss state with the electronic
ground state, respectively~\cite{Lippert:CPL376:40, Conde:CP530:25}. However, these studies suffered
from the overlapping signals of different cluster sizes described above. Therefore, product channels
could not be investigated and the long-time relaxation dynamics were dominated by contributions from
the fragmentation of the larger clusters.

Here, we present the results of an ultrafast-dynamics pump-probe experiment that utilised the
electrostatic deflector to produce a high-purity bimolecular solute-solvent aggregate sample, in
combination with velocity-map imaging (VMI) mass spectrometry to disentangle the reaction products.
We utilised these capabilities to study the UV-induced dynamics of the
prototypical \indolew complex including eventual dissociation on the
picosecond timescale. We demonstrate that our high-purity sample allows us to investigate product
channels and to follow the dynamics of \indolew on long timescales. This provides significant
additional insight into these prototypical reaction dynamics, \ie, it provides evidence for an
incomplete hydrogen-transfer process in \indolew. Overall, our experimental approach, with the
bimolecular reactants well-defined and the product channels clearly identifiable in real time,
propels us a major step forward to unravel the complete pathways in bimolecular reaction systems.

\section{Results}
\label{sec:results}
The ultrafast chemical dynamics of \indolew was investigated in a molecular-beam apparatus
containing an electrostatic deflector and a velocity-map-imaging ion detector, see
\autoref[c]{fig:yields}. Using the deflector, we spatially separated \indolew from the other species
in the molecular beam, \eg, \indole, \HHO, and helium seed gas~\cite{Trippel:PRA86:033202}. This
resulted in a high-purity \indolew sample, as shown in \supsec{1}. Most signal from a
small remaining fraction of \HHOd could be discriminated by ion-momentum imaging. UV pump pulses
with a central wavelength of 269~nm electronically excited \indolew. The reaction products were
detected through ionisation with a delayed NIR probe pulse with a central wavelength of 1320~nm and
velocity-map imaging of the generated ions. The high-purity sample provided by the deflector allowed
us to investigate all product channels in an effectively background-free manner, since the product
molecules, \ie, indole and water, are not present in the molecular beam in the interaction region --
whereas in traditional experiments they were present in the direct molecular beam in large amounts
that obscured these product signals. Thus the purified reactants allowed us to identify the
appearance of \indolep and \HHOp product ions that originate from \indolew.

\autoref[a,~b]{fig:yields} show the measured \indolewp (blue dots), \indolep (red diamonds), and
\HHOp (green squares) ion signals as a function of the pump-probe delay. The \HHOp and \indolewp
signals are scaled up for improved visibility by factors 10 and 5, respectively.
\autoref[a]{fig:yields} shows the long-time relaxation dynamics, whereas~\autoref[b]{fig:yields}
zooms in on the short delays for \indolewp and \indolep. Dynamic effects are clearly visible in all
three signals: All channels exhibit a constant ionisation signal for negative delays that increases
when the laser pulses temporally overlap. The \indolewp signal shows a fast increase followed by a
decay with a fast and a slow component, whereas the \indolep product signal contains a fast increase
followed by a slow increase. We could clearly observe a delay of the fast increase in the \indolep
signal compared to the fast increase in the \indolewp signal. The \HHOp product signal shows a slow
increase as a function of the pump-probe delay, see \autoref[a]{fig:yields}. On top of these general
dynamics, we observed oscillations with a period of 1.67~ps for \indolewp and \indolep, which are in
phase between the \indolewp and \indolep signals. High-temporal-resolution measurements for \indolep
at $\ordsim120$~ps showed that the oscillations are not damped on that timescale,
see Supplementary Figure 4. Our measurements did not reveal any oscillations in 
the \HHOp signal.

To disentangle the underlying dynamics of these time-dependent ion yields we employed a reaction
model, see \methods. The observations could be described well by a five-level model, allowing to
populate the levels sequentially:
\begin{equation}
   \begin{split} 
      \label{eq:model}
     &~ \indolew \: S_0 \: (1) \xrightarrow{\text{UV}} \indolew \: \pps \: (2) \\
     &~ \xrightarrow{\tau_2}  \indolew \: \pss \: (3) \\
     &~ \xrightarrow{\tau_3} \indolew \: S_0 \: (4) \\
     &~ \xrightarrow{\tau_4} \indole + \HHO \: (5)
   \end{split}
\end{equation}
with the time constants $\tau_{i}$ coupling states $i$ and $i+1$. The dynamics in
\indolean{n}~\cite{Lippert:PCCP6:2718} and \indolewn{n}~\cite{Lippert:CPL376:40} clusters were
previously described with similar models, albeit with six and four states. Here, we needed five
levels to accurately describe the dynamics we observed in \indolew. We started with all population
in the electronic ground state $S_0~(1)$ of \indolew and used Maxwell-Bloch equations to find the
time-dependent populations of all states. The ion yields are calculated as linear combinations of
these populations, which for \indolep and \indolewp were multiplied by a cosine function
representing the oscillations. We note that these timescales, esp.~$\tau_2$, do not necessarily
correspond to pure electronic interconversion processes, \emph{vide infra} and that the model does
not distinguish between the $^1L_b$ and $^1L_a$ electronic states, but rather considers them as one
state.

We fitted the coefficients for the linear combinations as well as the time constants to the
experimental data and obtained $\tau_2=445\pm71$~fs, $\tau_3=13\pm2$~ps, and $\tau_4=96\pm10$~ps.
These results were used to calculate the yields of \indolewp, \indolep, and \HHOp shown by the blue,
red, and green solid lines in \autoref[a,~b]{fig:yields}, respectively. \autoref[a]{fig:yields}
shows the signals computed from the Maxwell-Bloch equations directly, whereas
\autoref[b]{fig:yields} includes the cosine functions describing the oscillations. This model
matches the experimental data very well.
\begin{figure}
   \includegraphics[width=\linewidth]{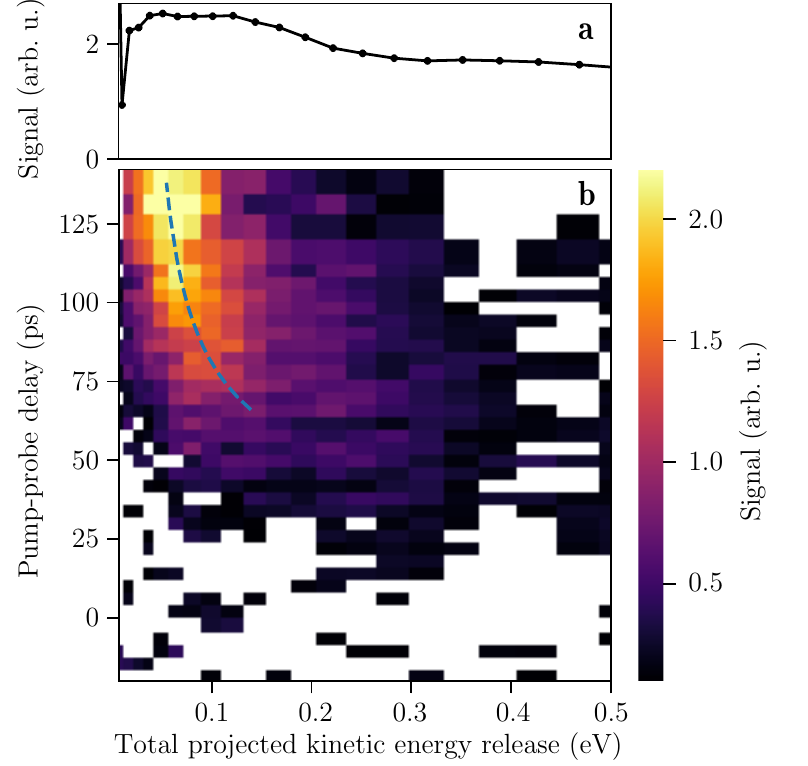}%
   \caption{\textbf{Time evolution of the total kinetic-energy release distribution of \HHOp.}
      \textbf{a}~Total kinetic-energy release (TKER) distribution for negative pump-probe delays,
      based on the projected kinetic-energy distribution of \HHOp. \textbf{b}~Time-dependent
      TKER-distribution differences from the neutral dissociation of \indolew after subtraction of
      the delay-independent signal. The blue dashed line represents the mean TKER obtained from an
      ion-dipole-interaction model, see~\methods.}
   \label{fig:H2O}
\end{figure}
Additional information about the underlying dynamics was obtained from the projected
kinetic-energies of \HHOp, \ie, the delay dependence of the total kinetic energy release (TKER) in
the neutral dissociation $\indw\longrightarrow\ind+\HHO$, see \methods. To separate the
delay-dependent signal from the constant \HHOp signal from dissociation of \HHOd, we first
determined the mean TKER distribution for negative pump-probe delays, when the \HHOp ion yield is
constant. This distribution, which is shown in~\autoref[a]{fig:H2O}, was subtracted from all TKER
resulting in the TKER changes shown in \autoref[b]{fig:H2O}. For short delays, the TKER distribution
is similar to the static background. However, a dynamical signal develops when the delay increases:
The mean TKER decreases and the distribution gets narrower.

\begin{figure*}
   \includegraphics[width=\linewidth]{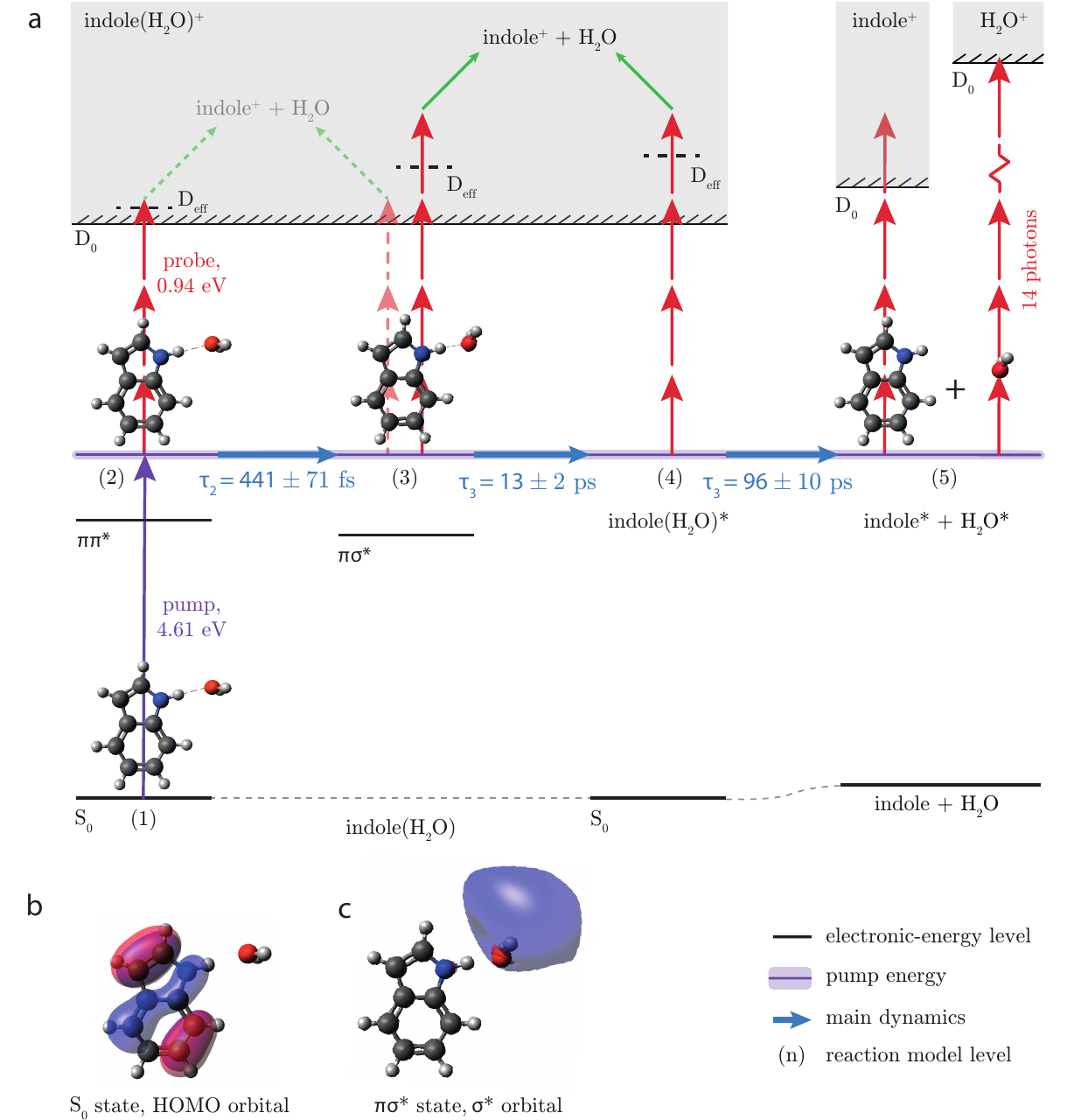}%
   \caption{\textbf{Schematic of the UV-induced photochemistry in \indolew.} a) The purple arrow
      indicates the UV pump-photon energy yielding the overall energy in the system after absorption
      of an UV photon, indicated by the purple line, which is conserved throughout the dynamics. The
      shaded purple area represents the corresponding bandwidth. Blue horizontal arrows depict the
      different steps in the observed photochemistry with corresponding timescales; see text for
      details. The red arrows indicate the absorption of NIR probe photons leading to ionisation.
      The black dashed lines labelled D$_{\text{eff}}$ represent the cationic states that need to be
      accessed to efficiently ionise from a particular initial state in \indolew; see text for
      details. The green arrows represent the dissociative ionisation process, leading to the
      formation of \indolep and \HHO. Dashed arrows indicate a significantly lower probability for a
      specific process to take place. The $\indole+\HHO$ product channel contains a lot of
      vibrational energy. Note that the energies in this figure are not to scale b) and c) Computed
      isosurface representations of the relevant molecular orbitals of \indolew. b) The highest
      occupied molecular orbital (HOMO), a $\pi$-orbital, for the equilibrium geometry in the $S_0$
      state. c) The $\sigma^*$ orbital in the equilibrium geometry of the \pss state.}
   \label{fig:interp}
\end{figure*}

\section{Discussion}
\label{subsec:discussion}
Based on our measurements and reaction model we expect the dynamics depicted
in~\autoref[a]{fig:interp} to occur. In short: The UV laser excites \indolew from the electronic
ground state $S_0~(1)$ to the electronically excited $\pps$ state~(2). The system then interconverts
to the optically dark $\pss$ state, which is accompanied by nuclear and electronic rearrangements
into (3). Subsequently, interconversion takes place to the vibrationally hot $S_0$ state~(4). This
vibrationally-hot electronic-ground-state \indolew dissociates into separate vibrationally-excited
\indole and \HHO molecules $(5)$, most likely \emph{via} so-called statistical unimolecular decay.
Moreover, vibronic-wavepacket dynamics occur in this system, which were taken into account in the
reaction model. In the following, the different steps will be discussed in more detail.

When the UV laser (purple arrow in~\autoref[a]{fig:interp}) excites \indolew to the \pps state, the
system can be ionised by the subsequent absorption of three NIR photons (left red arrows) through
the electronic ground state $D_0$ of the cation, which has a similar geometry as the \pps
state~\cite{Unterberg:JCP113:7945} and an ionisation energy of \Ei$=7.37$~eV~\cite{Braun:JPCA:3273}.
The vertical transition is indicated by the dashed horizontal line labelled $D_{\text{eff}}$ in
\autoref[a]{fig:interp}. This results in the fast initial increase in the \indolewp signal. Due to
the excess energy above the ionisation potential, fragmentation of \indolewp can
occur~\cite{Trippel:PRA86:033202}, leading to a small contribution to the fast increase in the
\indolep signal, see \supsec{4}.

The subsequent interconversion from the \pps state to the \pss state with $\tau_2=445\pm71$~fs is
accompanied by nuclear rearrangement and electron transfer, as shown by the results of \emph{ab
   initio} calculations that are described in \methods. In the equilibrium geometry of the \pss
state, depicted in \autoref[c]{fig:interp}, the \HHO molecule tilts along the in-plane-bending
coordinate of the cluster and the N-O distance decreases, whilst the N-H distance increases with
respect to the geometry in the $S_0$ state, depicted in \autoref[b]{fig:interp}. Moreover, an
electron is transferred from a $\pi$-orbital on the \indole moiety to a $\sigma^*$-orbital localised
around the \HHO molecule.
This is depicted by the calculated isosurfaces of the orbital amplitudes plotted in the structures
in \autoref[b,c]{fig:interp}.

In order to efficiently ionise the system from this geometrically different \pss state with a
different amount of vibrational energy (3), \ie, in a vertical transition, it seems that a
higher-energy cationic state needs to be accessed, depicted by the dashed horizontal line labelled
D$_{\text{eff}}$ in~\autoref[a]{fig:interp}. We suspect that this cationic state is above the
binding energy of \indolewp of 0.6~eV~\cite{Mons:JPCA103:9958}, resulting in more fragmentation of
the cation and correspondingly more \indolep signal~\cite{Mons:JPCA103:9958}. This process likely
required four NIR photons, see the \supsec{4} for more details. Overall, this leads to a
decrease in the \indolewp signal and a corresponding increase in the \indolep signal. This agrees
with the fast increase in the \indolep signal being delayed with respect to the fast increase in the
\indolewp signal.

We attribute the 1.67~ps oscillations overlayed on these electronic and nuclear dynamics to
wavepacket dynamics, for instance due to \indole ring breathing modes or C--H-stretch vibrations,
which modified the transition strengths in the system, see \supsec{5}. We expect that
these dynamics occurred due to the coherent excitation of vibrational states with an energy
difference of 20~\invcm, within the bandwidth of the pump laser depicted by the shaded purple area
in~\autoref[a]{fig:interp}, and that they manifest themselves as a modification of the
vertical-ionisation probability, which lead to the observation of these oscillations.

Subsequently, the \indolew complexes in the \pss state interconvert within $\tau_3=13\pm2$~ps to the
$S_0$ state through nonadiabatic coupling, as was described before~\cite{Lippert:thesis:2004}. We
expect that this interconversion leads to vibrationally-excited \indolew clusters in the electronic
ground state, which were ionised less likely due to the increased \Ei for a vertical transition --
depicted by the dashed horizontal line D$_{\text{eff}}$ state in~\autoref[a]{fig:interp} -- and
which results in \indolewp ions that subsequently dissociate, leading to \indolep signal. This would
explain the slow decrease in the \indolewp signal and the slow increase in the \indolep signal.

We believe that the vibrationally-hot \indolew subsequently dissociates on a timescale of
$\tau_4=96\pm10$~ps into separate neutral \indole and \HHO molecules containing significant
vibrational energy. Both products could be ionised, which explains the increasing \indolep and \HHOp
signals. At least four NIR photons are needed to reach the ionisation energy for the
vibrationally-hot \indole, whereas a minimum of 14 NIR photons is needed for the ionisation of
ground-state \HHO.

An alternative pathway for the formation of \HHO molecules would be direct dissociation in the \pss
state. However, our measured TKER distributions, \autoref{fig:H2O}, match Maxwell-Boltzmann
distributions for delays $\larger65$~ps, which strongly indicates that the \HHO molecules are formed
in a statistical unimolecular-decay process in the $S_0$ state instead of a direct dissociation
process in the \pss state, see~\methods. Direct dissociation from the \pps state is in principle
possible but, also because of the short sub-picosecond lifetime of this state, irrelevant.

The mean of the TKER distributions changes as a function of the delay, see~\autoref{fig:H2O}. This
time evolution is explained by considering the ion-dipole interaction: When the \HHO molecule is
ionised and the separate \indole and \HHOp molecules are still relatively close, they will repel
each other due to the repulsion of the \HHOp charge and the \indole dipole, leading to an increase
in the kinetic energy. For longer delays, the increased distance between the two moieties results in
a correspondingly lower kinetic energy. The width of the TKER distribution is governed by the
wavepacket evolution along the ion-dipole interaction potential and decreases for longer delays. We
used a classical ion-dipole interaction model to describe the evolution of the TKER; details are
provided in~\methods. This yielded the dashed blue line in \autoref{fig:H2O}, which is in very good
agreement with our experimental results, indicating that the two moieties, on average, slowly move
apart with a speed $v\approx12$~m/s. This is another confirmation for the statistical
unimolecular-decay process in the $S_0$ state.

In order to shed light on the possible net hydrogen-transfer process taking place in the \pss state
of \indolew~\cite{Sobolewski:PCCP4:1093}, we investigated the \HHHOp signal. However, we did not
find any delay-dependent effect in the \HHHOp signal coming from neutral dissociation of \indolew.
This indicates that, although the N--H bond is stretched in the \pss state, the electron transfer is
not followed by proton transfer to \HHO for dissociation. We conclude that after UV excitation at
4.61~eV the \indolew aggregate either survives or dissociates into the individual \indole and \HHO
molecules, which themselves stay intact and slowly move away from each other.

Our pump-probe experiments on pure \indolew samples in combination with our reaction model based on
a five-level system provide new insight into the ultrafast processes occurring in this prototypical
solvated-biomolecule system, especially regarding the combined electronic and nuclear dissociation
dynamics and its product channels. The time constant $\tau_2=445\pm71$~fs that we assigned to the
interconversion from the \pps to the \pss state and accompanied nuclear rearrangement and electron
transfer is within the broad range $\tau_2=\range{150}{500}$~fs previously tentatively attributed to
dynamics on the \pss surface~\cite{Conde:CP530:25}. We note that $\tau_2$ in the present study is on
the same order as the instrument response function (IRF). We could not resolve the fastest time
constant of $\ordsim50$~fs~\cite{Lippert:CPL376:40} or $\range{20}{100}$~fs~\cite{Conde:CP530:25}
observed before, since it is shorter than the IRF. This time constant was previously assigned to
internal conversion from the \pps to the \pss state~\cite{Lippert:CPL376:40} or from the $^1L_a$ to
the $^1L_b$ state~\cite{Conde:CP530:25}, which are both \pps states. Regarding the time constant
$\tau_3=13\pm2$~ps, our interpretation is consistent with previous tentative assignments of a 14~ps
decay constant to interconversion from the \pss state to the $S_0$ state~\cite{Lippert:CPL376:40}.
While previous work was not able to distinguish products from different cluster sizes, our
well-defined-reactant study fully supports this finding. Furthermore, the $\tau_4=96\pm10$~ps time
constant we obtained for the bond breaking and formation of \HHO is comparable to the ones found for
the production of (NH$_3$)$_{n-1}$NH$_4^+$ resulting from dissociation of \indolean{n} clusters with
$n\leq2$~\cite{Lippert:PCCP6:2718}.

In conclusion, we demonstrated that our experimental approach using species-selected samples
to perform pump-probe studies of ultrafast chemical dynamics provides unprecedented details on
intermediates and reaction products and thus the chemical reaction dynamics. Creating a high-purity
sample of the prototypical solvated biomolecule \indolew enabled us to investigate its UV-induced
dissociation dynamics in intimate detail, well beyond previous experimental studies. We observed an
initial delay in the appearance of \indolep ions, which we ascribed to the ionisation of \indolew
from the \pps and \pss states resulting in different cationic states of \indolew with distinct
fragmentation probabilities. Moreover, we could follow the long-time relaxation dynamics in the
reaction products, which revealed clear evidence for an incomplete hydrogen-transfer process and
thus indicates that the indole chromophore is protected by the attached water against UV-induced
radiation damage. This is opposite to earlier theoretical predictions \cite{Sobolewski:PCCP4:1093},
but fully in line with previous experiments \cite{Lippert:CPL376:40}.

While in previous experiments the \HHOp product signal was completely obscured by the unavoidable
large amount of \HHO in the molecular beam, our purified samples allowed us, for the first time, to
experimentally determine the time constant of the hydrogen-bond-breaking process to
$\tau_4=96\pm10$~ps. Moreover, based on the kinetic-energy distributions of the \HHO products we
conclude that this biochemically important process~\cite{Tarek:PRL88:138101} occurs \emph{via}
statistical unimolecular decay in the electronic ground state. As ultrafast excited-electronic-state
deactivation after the absorption of UV photons could be essential for the photostability of
proteins~\cite{Domke:NatChem5:1755}, our results demonstrate how such mitigation of UV-induced
radiation damage through solvent interactions works.

Overall, these results demonstrate that our experimental approach combining the deflector,
velocity-map imaging, and pump-probe ultrafast time-resolved spectroscopy enables the observation of
complete chemical-reactivity pathways in chemical reactions of complex molecular systems. The
investigation of a bimolecular half-collision reaction allowed for the precise triggering of the
dynamics and presents a promising approach for more complex chemical systems.

Our approach can directly be combined with shorter, \ie, few-femtosecond or attosecond, laser pulses
and with tunable wavelengths to follow the energy-dependent ultrafast electronic and chemical
processes in complex reaction systems. This could be further aided by coincidence
measurements~\cite{Ullrich:RPP66:1463,Johny:protection:inprep}. On the other hand, high-energy UV
photons could be used in order to ionise the complex and its fragments with a single photon.
Moreover, diffractive imaging of the nuclear dynamics~\cite{Blaga:Nature483:194,
   Kuepper:PRL112:083002, Yang:Science361:64} would provide complementary detailed information on
the actual atomic structural dynamics. For instance, photoelectron-momentum-imaging and
laser-induced-electron-diffraction experiments would be a direct extension of the current
experiments and initial experiments are ongoing~\cite{Karamatskos:JCP150:244301, Wiese:PRR3:013089}.
These approaches should reveal the complete reaction pathway of the nuclear and electronic dynamics
that occur in these reactions. Such detailed insights will ultimately yield a deep understanding of
the formation and breaking of bonds and allow to develop a truly dynamical basis of chemistry.

\section*{Methods}
\label{sec:methods}
\subsection*{Experimental setup}
\label{subsec:setup}
The experimental setup~\cite{Trippel:MP111:1738, Trabattoni:NatComm11:2546} is shown schematically
in~\autoref[c]{fig:yields}. 95~bar of helium was bubbled through room-temperature water before
passing through the sample reservoir of an Even-Lavie valve containing \indole (Sigma-Aldrich,
$\geq99~\%$). The valve was operated at \celsius{110} and a repetition rate of 250~Hz. After passing
through two skimmers, the beam travelled through a 15.4~cm long electrostatic
deflector~\cite{Chang:IRPC34:557}. Applying a voltage of 20~kV between the electrodes of the
deflector created inhomogeneous electric fields that spatially dispersed and separated \indole and
\indolew based on their Stark effect~\cite{Trippel:PRA86:033202, Chang:IRPC34:557}, see the
\supsec{1}. Passing through another skimmer the molecules were intersected by the focused pump and
probe laser beams in the centre of a VMI spectrometer.

Indole(H$_2$O) was electronically excited by ultraviolet-light (UV) pulses with a central wavelength
of 269~nm, a pulse duration of $\ordsim650$~fs (full width at half maximum, FWHM), and a peak
intensity of $\ordsim2\cdot10^9$~\Wpcmcm. Near-infrared (NIR) pulses centred around 1320~nm with a
duration of $\ordsim70$~fs (FWHM) and a peak intensity of $\ordsim1\cdot10^{13}$~\Wpcmcm were used
to ionise the complex and its fragments. We used a wavelength of 1320~nm to avoid the three-photon
resonant excitation of \indole and \indolew at 800~nm and to have a similar ionisation step as in
previous work on UV-induced dynamics in \indolewn{n} clusters~\cite{Conde:CP530:25}. Both laser
beams were linearly polarised along $Y$, \ie, parallel to the detector plane, and focused into the
centre of the molecular beam with $4\sigma\approx90$~\um and $4\sigma\approx60$~\um for the
intensities of the UV and NIR beams, respectively. Unless mentioned otherwise, we adjusted the laser
powers such that the ionisation signal of \indolew with either of the two beams alone was
negligible.

The generated ions were accelerated towards a multichannel-plate and phosphor-screen detector using
the VMI spectrometer. Images were recorded with a CMOS camera at 500~Hz, alternating between
molecular-beam signal and background frames. The MCP was temporally gated in order to record ion
images for individual mass-to-charge ratios, which we refer to as mass-gated images. Indole$^+$ and
\indolewp signals were measured in a detuned-VMI mode to avoid saturation and damage of the central
part of the detector. \HHOp ions were measured in VMI mode at an increased NIR intensity of
$\ordsim1\cdot10^{14}~\Wpcmcm$, accounting for the relatively high ionisation energy
$\Ei\approx12.6$~eV of \HHO.

The delay between pump and probe laser pulses was scanned back and forth multiple times using a
motorised translation stage. Indole(H$_2$O)$^+$ and \indolep signals were recorded in the same
measurements, by scanning the pump-probe delay and the mass-gating simultaneously. We used a step
size of 417~fs for $t=\range{-1.535}{6.805}$~ps, made one step of 834~fs, and used a step size of
1.668~ps for $t>7.639$~ps. We used $\ordsim8\,000$ laser shots per data point. The \HHOp signal was
recorded for $t=\range{-5.705}{140.245}$~ps with a step size of 4.17~ps and $\ordsim41\,000$ laser
shots per data point. Data at 119.395, 123.565, and 140.245 ps was discarded due to instabilities in
the spatial overlap of the UV and NIR beams for the longest delays. The delays $t > 114$~ps do not
significantly alter the fit, but show that the \HHOp signal starts to level off. To improve the
statistics for \autoref{fig:H2O}, this data was combined with a second measurement for which the
\HHOp signal was recorded for $t=-19.391\ldots122.390$~ps with a step size of 1.668~ps and
$\ordsim340\,000$ laser shots per data point. The two data sets were merged and averaged over 4~ps,
and the data points at 118, 122, and 142~ps were discarded due to the non-perfect spatial overlap of
the laser beams.

\subsection*{Reaction model}
\label{subsec:kinmod}
We described the time-dependent ion signals using a reaction model for a five-level system,
see~\eqref{eq:model}. The Maxwell-Bloch equations that describe the evolution of the populations
$\rhos{i}{i}$ of state $i$ corresponding to this model are given by~\cite{Hertel:RPP69:1897,
   Lippert:thesis:2004}
\begin{equation}
  \label{eq:OBE}
  \begin{split}
     \rhosdot{1}{1} &= \frac{i}{2}\Omega_0 \, g(t) \, (\rhos{1}{2}-\rhos{2}{1}) \\
     \rhosdot{2}{2} &= -\frac{i}{2}\Omega_0 \, g(t) \, (\rhos{1}{2}-\rhos{2}{1})-\Gamma_{22}\rhos{2}{2} \\
     \rhosdot{2}{1} &= -\frac{i}{2}\Omega_0 \, g(t) \, (\rhos{1}{1}-\rhos{2}{2})-(\Gamma_{21}-i\Delta\omega)\rhos{2}{1} \\
     \rhosdot{1}{2} &= \frac{i}{2}\Omega_0 \, g(t)^* \, (\rhos{1}{1}-\rhos{2}{2})-(\Gamma_{21}+i\Delta\omega)\rhos{1}{2} \\
     \rhosdot{3}{3} &= \Gamma_{22}\rhos{2}{2} - \Gamma_{33}\rhos{3}{3} \\
     \rhosdot{4}{4} &= \Gamma_{33}\rhos{3}{3} -\Gamma_{44}\rhos{4}{4} \\
     \rhosdot{5}{5} &= \Gamma_{44}\rhos{4}{4},
  \end{split}
\end{equation}
where $\Gamma_{ii}=1/\tau_{i}$ and $g(t)=\exp({-\frac{1}{2}(t/\tau_{\text{IRF}})^2})$, which
represents the instrument-response function (IRF) with $\tau_{\text{IRF}}=381$~fs, see
\supsec{3}. We assumed $\Delta\omega=0$ and
$\Gamma_{21}=\Gamma_{22}/2$~\cite{Hertel:RPP69:1897}. The Rabi frequency was estimated to
$\Omega_0=3.4~\text{ps}^{-1}$ based on the peak intensity and duration of the UV pulses and an
estimated transition dipole moment $\mu_{12}\approx15~e$\,pm~\cite{Lippert:thesis:2004}.

Initially, all population was in state~1, \ie, \mbox{$\rhos{1}{1}(-\infty)=1$}. Integrating
\eqref{eq:OBE} yields the delay-dependent populations of the different states. The simulated
ion-signal intensities for \indolewp, \indolep, and \HHOp are given by linear combinations of the
different populations $\rhos{i}{i}$:
\begin{equation}
  \label{eq:ionyields}
  \begin{split}
     I_{\indolewp}(t) &= p^\text{osc}_{\indolewp}(t) \, \sum_{i=2}^{4} A_i\rhos{i}{i}'(t)  \\
     I_{\indolep}(t) &= p^\text{osc}_{\indolep}(t) \, \sum_{i=2}^{5} B_i\rhos{i}{i}'(t) \\
     I_{\HHOp} &= C\rhos{5}{5}',
  \end{split}
\end{equation}
with the population $\rhos{i}{i}'(t)=f(t)\otimes\rhos{i}{i}(t)$ of state $i$ after convolution with
a Gaussian function $f(t)$ with a FWHM of 70~fs, which represents the intensity envelope of the NIR
pulse~\cite{Hertel:RPP69:1897}. The decay constants $\tau_2$, $\tau_3$, and $\tau_4$ as well as the
coefficients $A_i,B_i$ and $C$ in \eqref{eq:ionyields} were fit using a Levenberg-Marquardt
algorithm and a reduced-$\chi^2$ objective function of simulated against background-corrected
experimental ion signals, \autoref[a]{fig:yields}. The oscillations in the \indolep and \indolewp
signals were modelled as $p^\text{osc}_j(t)=a_j+b_j\cos(\omega{t}+\phi)$, with $j=\indolewp$ or
\indolep. These parameters were optimised once using a high-resolution measurement of the
oscillations and then fixed in the fitting procedure to $a_{\indolewp}=1.04$, $b_{\indolewp}=0.14$,
$a_{\indolep}=0.92$, $b_{\indolep}=0.11$, $\omega=2\pi\cdot0.60~\text{THz}=3.77~\text{THz}$ and
$\phi=1.77$.

The best fit yielded $\tau_2=445\pm71$~fs, $\tau_3=13\pm2$~ps, and $\tau_4=96\pm10$~ps with a
reduced $\chi^2$ of $\chi_\nu^2=1.42$ and a coefficient of determination $R^2=0.999$. The resulting
time-dependent contributions of the individual states to the ion signals are shown in Figure~S3 in
the \supsec{4}.

\subsubsection*{Evolution of the total kinetic energy release}
\label{subsec:TKER}
For the low-kinetic-energy \HHOp signal resulting from neutral dissociation, we investigated the
kinetic-energy distributions as a function of the pump-probe delay. The total projected kinetic
energy release (TKER) assuming neutral dissociation can be computed using
\begin{equation}
   \text{TKER} = E^K_{\HHO}\frac{m_{\indolew}}{m_{\indole}},
\end{equation}
where $E^K_{\HHO}$ is the kinetic energy (KE) of \HHO and $m_{\indole}$ and $m_{\indolew}$ are the
masses of \indole and \indolew, respectively.

\begin{figure}
   \includegraphics[width=\linewidth]{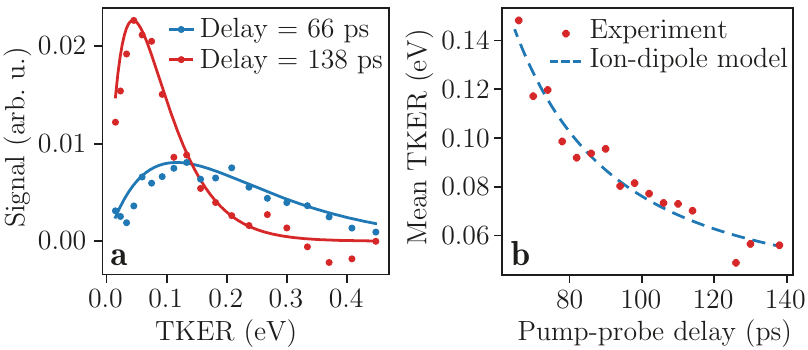}%
   \caption{\textbf{Evolution of the total projected kinetic energy release.} \textbf{a,} TKER
      distributions for pump-probe delays of 66~ps (blue) and 138~ps (red). The dots are the
      experimental distributions after subtracting the mean TKER distribution for negative
      pump-probe delays, whereas the solid lines show fitted statistical distributions. \textbf{b,}
      Mean TKER values obtained from the experimental distributions (red dots) and a fit through
      these data points (blue dashed line) based on an ion-dipole interaction model, see the text
      for details.}
   \label{fig:TKER}
\end{figure}

\autoref{fig:H2O} shows how the projected TKER distributions evolve as a function of the pump-probe
delay. The blue and red dots in \autoref[a]{fig:TKER} represent these distributions for delays of
66~ps and 138~ps, respectively. We subtracted the mean TKER distribution for negative pump-probe
delays, shown in \autoref[a]{fig:H2O}, which represents the delay-independent \HHOp signal. We note
that the recorded projected KEs are, in principle, different from the real three-dimensional KEs,
but in line with the geometric orientation due to the dipole excitation and the clear rings in the
VMIs we assume the projected KEs are a reasonable approximation for the following discussion.

We successfully fitted Maxwell-Boltzmann distributions to all experimental velocity distributions
for delays $\larger65$~ps. The resulting experimental KE spectra and statistical distributions for
delays of 66~ps and 138~ps are shown in~\autoref[a]{fig:TKER}. This indicates that the \HHO
molecules resulted from statistical unimolecular decay, in the vibrationally-hot electronic ground
state, instead of direct dissociation from an electronically excited state, \eg, the \pss
state~\cite{Godfrey:PCCP17:25197}.

From the data in \autoref{fig:H2O} it is clear that the mean TKER decreases, \ie, the mean kinetic
energy of the \HHO molecules, decreases as a function of the delay. The distribution also becomes
narrower for longer delays, in line with a statistical process in which the hottest reactants
dissociate the fastest and earliest. To visualise the temporal evolution of the TKER, the mean
values of the fitted statistical distributions as a function of the pump-probe delay are shown in
\autoref[b]{fig:TKER}.
This temporal evolution of the TKER distribution could be explained by an ion-dipole interaction
model, detailed below, based on previous descriptions of similar Coulomb-repulsion effects, \eg, in
the dissociation dynamics of small stable molecules~\cite{Baklanov:JCP129:214306,
   Erk:Science345:288, Boll:StructDyn3:043207}.

When a vibrationally-hot \indolew molecule dissociates and is afterwards ionised by the probe laser
an electron can in first instance tunnel back and forth between the \indole and \HHO moieties.
However, when the distance between the two molecules increases, the electron at some point localises
on one of the two monomers. If the hole localises on \HHO we get $\indole+\HHOp$ and can detect the
\HHOp ion. Since the \indole side pointing towards the \HHOp ion is the positive end of the indole
dipole moment in the $S_0$ state~\cite{Kang:JCP122:174301}, the two moieties will repel each other
due to the ion-dipole interaction. The interaction energy $V$ between \indole and \HHOp, neglecting
higher-order terms such as the ion-induced dipole and dipole-dipole interactions, is given by
\begin{equation}
\label{eq:Viondip}
V(t) = \frac{q_{\text{ion\/}} \, \mu\cos\theta}{4\pi\epsilon_0R(t)^2},
\end{equation}
with the charge $q_{\text{ion}}$ of the \HHO molecule, \ie, $q_{\text{ion}}=1$, the dipole moment
$\mu$ of the \indole molecule, the interaction angle $\theta$, the vacuum permittivity $\epsilon_0$,
and the distance $R$ between the ion and the dipole. We assumed that $\mu=1.96$~D, \ie, the dipole
moment of \indole in the $S_0,v=0$ state~\cite{Kang:JCP122:174301}, and that $\theta$ is
time-independent.

We further assumed that the two fragments dissociate with a constant velocity $v$ starting from an
equilibrium distance $R_{\text{eq}}$ at a delay $t_{\text{d}}$. We used the following expression for
the time-dependent ion-dipole distance $R(t)$
\begin{equation}
R(t) = R_{\text{eq}} + v(t-t_{\text{d}}),
\end{equation}
which we substituted in \eqref{eq:Viondip}. The total kinetic energy in the system is now given by:
\begin{equation}
\text{TKER}(t) = E_{\text{pump}} + V(t) - E_{\text{a}},
\end{equation}
with $E_{\text{pump}}$ the energy of a pump-laser photon, 4.61~eV, and $E_{\text{a}}$ the asymptotic
internal energy of the two fragments, \ie, for $R=\infty$.

Fitting $R_{\text{eq}}$, $v$, $t_{\text{d}}$, $\theta$ and $E_{\text{a}}$ to the experimental
evolution of the mean TKER yielded the blue dashed line in~\autoref{fig:H2O} and
\autoref[b]{fig:TKER}. We found that $R_{\text{eq}} = 490$~pm, $v=12$~m/s, $t_{\text{d}}=51$~ps,
$\theta=38^\circ$, and $E_{\text{a}}=4.57$~eV. The fit reproduces the experiment well, which
indicates that the ion-dipole interaction drives the observed evolution of the TKER. The low
asymptotic TKER, \ie, $E_{\text{pump}}-E_a$, which is below 50 meV, is an additional indication for
the \HHO molecules resulting from dissociation in the $S_0$ state instead of the repulsive \pss
state, since such repulsive states generally lead to a higher asymptotic energy.

We note that our simple model neglects attractive interactions such as the ion-induced dipole
interaction, which could possibly be relevant due to the relatively high polarisability of indole.
Moreover, $\mu$ could change significantly due to vibrational excitation. However, our experimental
data does not contain sufficient information to accurately take this into account and corresponding
literature values were not available. As these effects partly cancel each other and based on the
very good agreement of our model and experimental data, our results still provide clear insight into
the origin of the TKER of \HHOp signal and, correspondingly, the actual dissociation dynamics of
\indolew.

\subsection*{\emph{Ab initio} calculations}
\label{subsec:abinitio}
We performed \emph{ab initio} calculations on \indolew using
P\textsc{si}4~\cite{Parrish:JCTC13:3185}. The ground state geometry was optimised using
density-fitted second-order M{\o}ller-Plesset perturbation theory (DF-MP2) calculations using an
aug-cc-pVTZ basis set. We used equation-of-motion coupled-cluster singles and doubles (EOM-CCSD) in
combination with an aug-cc-pVDZ basis set to optimise the geometry in the \pss state.

\section*{Data availability}
The data that support the findings of this study are available from the repository at
\url{https://doi.org/10.5281/zenodo.7024411}.

\section*{Code availability}
The script used to solve the Maxwelll-Bloch equations is available from the repository at
\url{https://doi.org/10.5281/zenodo.7024411}.

\section*{Acknowledgements}
We thank Joss Wiese, Melby Johny, and Oriol Vendrell for fruitful discussions and Jovana Petrovic
and Terry Mullins for support of the experiments.

We acknowledge financial support by Deutsches Elektronen-Synchrotron DESY, a member of the Helmholtz
Association (HGF), also for the provision of experimental facilities and for the use of the Maxwell
computational resources operated at DESY. This work has been supported by the Clusters of Excellence
``Center for Ultrafast Imaging'' (CUI, EXC~1074, ID~194651731) and ``Advanced Imaging of Matter''
(AIM, EXC~2056, ID~390715994) of the Deutsche Forschungsgemeinschaft (DFG) and by the European
Research Council under the European Union's Seventh Framework Program (FP7/2007-2013) through the
Consolidator Grant COMOTION (614507). J.O.\ gratefully acknowledges a fellowship by the Alexander
von Humboldt Foundation.

\section*{Author contributions}
J.K.\ conceived the experiments and supervised the study. J.O.\ performed the experiment with
contributions from S.T. J.O.\ analysed the data. All authors were involved in interpreting the data
and discussing the results. J.O.\ and J.K.\ wrote the manuscript with contributions from S.T.

\section*{Competing interests}
The authors declare no competing interests.

\onecolumngrid
\clearpage

\end{document}